\documentclass[12pt]{article}
\begin{document}

\title{Vacuum instability and gravitational collapse}
\author{Michael Hewitt \\
Computing Services, Canterbury Christ Church University College,\\
North Holmes Road, Canterbury, CT1 1QU, U.K. \\ email:
mike@cant.ac.uk,
 tel:+44(0)1227-767700.}
\date{11 December 2002}
\maketitle
\begin{abstract}
We argue that $(0,1)$ heterotic string models with 4 non-heterotic
spacetime dimensions may provide an instability of the vacuum to
gravitational polarization in globally strong gravitational
fields. This instability would be triggered during gravitational
collapse by a non-local quantum switching process, and lead to the
formation of regions of a high temperature broken symmetry phase
which would be the equivalent in these string models of
conventional black holes. \

\

PACS numbers:  11.25.Sq, 11.25.Mj, 04.70Dy.

\end{abstract}

\section{Introduction}

\ \ \ In this paper we will argue that in heterotic string theory
[1] with 4 spacetime dimensions in non-heterotic form (i.e that
can be bosonised without twisted boundary conditions, see [2]),
the equivalent of conventional black holes are regions of a high
temperature broken symmetry phase characterised by a condensate of
strings. These would be the ultimately compressed states formed in
gravitational collapses unchecked in any other way. The
thermodynamic properties of these regions would closely match
those of the corresponding Hawking black holes [3].
 The analysis presented here has its origin in an attempt to make
sense of the interacting string spectrum in the context of thermal
duality [4]. It is intended that this paper will give an improved
analysis both of the nature of the final state and of the
nucleation mechanisms that can lead to the formation of such
states, remedying some of the shortcomings of the scenario
proposed in [4]. In section 2 we treat the static case and the
region of string condensate equivalent to a black hole, and in
section 3 we discuss a nucleation mechanism that could lead to the
formation of such regions in the weak tidal gravity of a large
collapsing aggregate of matter. Our approach differs from that of
Leonard Susskind [5,6] who argues that the relationship between
strings and black holes may be understood without introducing a
modification of the gravitational field structure around black
holes. The scenario outlined below would lead to the formation of
a new kind of object which we may call `string stars' in extreme
astrophysical collapses. The formation such an object would take
place through a non-local quantum effect producing a phase
transition. The argument below does not demonstrate conclusively
that such a process actually takes place - rather, the intention
is to alert the reader to the possibility that such a mechanism
may be latent within string theory. \

\section{Broken Symmetry Phase}
\ \ \ At the Hagedorn temperature the partition sum $Z$ for the
spectrum of a free string diverges [7]. This is associated with
the appearance of a tachyon [8,9] in the string spectrum which we
will call the thermalon $\phi$. For free strings, this appears as
a pole in the one loop diagram. This pole represents the
contribution of long string loops propagating around the thermal
circumference $\beta$ in imaginary time, and by modular invariance
this has a dual interpretation as the contribution of thermalons
propagating around the long loops represented by the corresponding
string configurations. In this picture, the thermalon summarises
the thermal behaviour of the whole string spectrum. This suggests
that the condensation of strings at the Hagedorn temperature, when
quartic self interactions are included, would be reminiscent of
the condensation of Higgs bosons, and that a high temperature
broken symmetry phase should exist [10]. Although the thermalon
itself exists only in the Euclidean regime, on continuing
amplitudes back to Minkowski spacetime the contribution of the
thermalon represents the effects of condensed strings in the high
temperature phase. Now it would appear that to produce a region of
this high temperature phase would be very difficult - a very large
amount of energy would need to be expended to bring thermal
radiation in the region up to the temperature where long strings
would condense. However, we will argue that the required
temperatures could be realised by geometric thermal effects in a
strong gravitational field. An observer suspended against the
strong gravitational acceleration $\alpha$ near an event horizon
would experience a heat bath of temperature $2\pi/\alpha$ [11],
while the energy density around a black hole of mass $M$ is
$O(1/M^4)$ [12], which is small and independent of $\alpha$. As
this geometric temperature reaches the Hagedorn point, the
observer's interaction with the heat bath could have the effect of
precipitating the condensation transition.  Thus, where the
geometric temperature of a static observer is at or above the
transition temperature, the usual vacuum state may be modified by
a symmetry breaking phase transition, and the difference $\Delta
\ln{Z}$ per unit volume between the two phases will have a finite,
non-zero value. The broken symmetry phase (BSP) exists between
$\beta = \pi / (1+\sqrt{2}) $, and $\beta = \pi (1+\sqrt{2}) $,
which we will call the upper and lower Hagedorn points
$\beta_{+}$, $\beta_{-}$ respectively.
 The formula for the free energy as a function of $\phi$, the thermalon
scalar field and $\beta$ in quartic approximation is [10]:
\begin{equation}
F = (-6 + \frac{\pi^{2}}{\beta^{2}} + \frac{\beta^{2}}{\pi^{2}})
\phi^{\ast}\phi + \lambda \phi^{\ast 2}\phi^{2} \label{QF}
\end{equation}
where $\lambda \sim g_{4}^{2}$. This depends upon $\beta$ through
the combination
\begin{equation}
\gamma^{2} = (\beta/ \pi)^{2}+ (\beta / \pi)^{-2} \label{GAM}
\end{equation}
which is invariant under the duality transformation
\begin{equation}
\beta \rightarrow \pi^{2} \beta^{-1},
\end{equation}
as a consequence of a compactification duality of imaginary time.
Strings at finite temperature have a lattice of heterotic winding
numbers in imaginary time, where the winding numbers $(a,b)$
represent null coordinates of a point in an even self dual
Lorentzian lattice. These winding numbers give a conformal weight
of
\begin{equation}
a^2\pi^{-2}\beta^2 + b^2\pi^{2}\beta^{-2}
\end{equation}
The thermalon itself has winding numbers $(1,1)$ in the heterotic
string theory [13] and so is invariant up to a phase under the
thermal duality arising from modular invariance:
\begin{equation}
a \rightarrow b, b \rightarrow a, \beta \rightarrow \pi^{2}/\beta
\end{equation}
which is realised as a reflection in heterotic imaginary time. A
change of metric $g_{00} \rightarrow \exp{(2\tau)} g_{00}$ is
represented by the Lorentzian transformation
\begin{equation}
\beta \rightarrow \exp{(\tau)} \beta. \label{TAU}
\end{equation}

From the viewpoint of quantum geometry, the effective
circumference of imaginary time for the thermalon, is $\gamma$.
The overall physics of the BSP however will not be dual symmetric
as spacetime is described by the level zero field $g_{\mu \nu}$
and the coupling of $g_{\mu \nu}$ to $F$ is not dual symmetric.
The contribution to $\ln{Z}$ from the thermalon in the BSP is
\begin{equation}
\Delta \ln{Z}(\beta, \phi) = -\beta \Delta F.
\end{equation}
The difference in energy density produced by the high temperature
phase is given by
\begin{equation}
\Delta \rho = -\frac{\partial}{\partial \beta} \Delta \ln{Z}
\label{DR}
\end{equation}
obtained by varying $g_{00}$. We may note that in thermal
equilibrium the value of $\beta$ will vary from place to place
according to
\begin{equation}
\sqrt{g_{oo}} \beta = \mathrm{const}. \label{VB}
\end{equation}
We might expect that the change in entropy density due to the
phase transition is given by
\begin{equation}
\Delta S = (1-\frac{\partial}{\partial\ln{\beta}})\Delta\ln{Z}
=\beta^{2} \frac{\partial \Delta F}{\partial \beta} \label{S0}
\end{equation}
i.e.
\begin{equation}
\Delta S = \beta (\Delta \rho - \Delta F) \label{S1}
\end{equation}
 However, a modification of this formula is needed due to the
conservation properties of energy in a gravitational field. The
essential property of eq.(\ref{S1}) is that because usually
\begin{displaymath}
\int \beta \rho dv = \mathrm{const}
\end{displaymath}
where $dv$ is the 3 dimensional volume element, we have $\int
\Delta S dv$ maximised when $\int \Delta F dv$ is minimised.
 The point which we now want to make is that the integral of the
gravitational source density $\epsilon = \rho + \mathrm{Tr}(p) =
T_{00}+T_{11}+T_{22}+T_{33}$ is the quantity which must be
constrained to remain the same to give the same external
gravitational field. For example, the Schwarzschild metric for
mass $M$ may be truncated to a flat interior at a surface where
the gravitational acceleration is $O(1)$ in string tension units.
If the transition is concentrated in a region with thickness
$O(1)$, the transverse pressure will be $O(1)$ but the total local
energy will only be $O(M)$, or $O(1)$ as viewed by a distant
observer. (The energy is supported against gravity by the pressure
as at the top of a dome.) The gravitational source in this case is
almost entirely converted to pressure. The distinction between
$\epsilon$ and $\rho$ only matters if the pressure is contained by
gravity - otherwise the flow of momentum is returned by the
containment vessel and the integral of $\beta \mathrm{Tr}(p)dv$
vanishes. Thus we require
\begin{equation}
\int \beta \epsilon dv = \mathrm{const} = M. \label{C}
\end{equation}
and therefore that $\int \Delta S dv$ should be maximised when
$\int \beta \Delta F dv$ is minimised subject to eq.(\ref{C}). If
we write
\begin{equation}
\Delta \epsilon = \Delta F + \Delta S/\beta \label{DE}
\end{equation}
so that $\Delta S \equiv \beta (\Delta \epsilon - \Delta F)$,
replacing eq.(\ref{S1}), then $\Delta S$ has the required
property. Note that eq.(\ref{DE}) is consistent with the
degeneracy of states needed with a Boltzmann factor based on
$\epsilon$ rather than $\rho$, i.e.
\begin{equation}
Z = \sum \exp{-\int \beta \epsilon dv}
\end{equation}
as would be expected for the statistical distribution of a
conserved quantity.

Now $\Delta S$ represents the difference in entropy between the
BSP and the conventional vacuum state. Thus $\ln{\Delta S}$
represents the effective degeneracy of states in the BSP with
$<\phi> \neq 0$. The origin of this degeneracy will be discussed
below. Now $\Delta F$ has a minimum at $\beta = \pi$, so $\Delta
\ln{Z(\beta, \phi)} = -\beta \Delta F(\beta, \phi)$ will have a
maximum at some $\beta_{P}(\phi)$ with $\pi < \beta_{P}(\phi) <
\beta_{-}$.
 For a region in which $\phi$ is approximately constant,
terms involving $\partial_{a}\phi$ may be neglected. Now a
space-filling solution with $\phi$ and $\beta$ constant should
exist by an application of the mean value principle. As $\beta$
varies, let $\phi_{m}(\beta)$ be the circle of values for which
$\Delta \ln{Z(\beta,\phi)}$ is maximised with $\phi$ constant in
space and time. Now let $\beta_{M}$ be the value of $\beta$ for
which $\Delta \ln{Z(\beta,\phi_{m}(\beta))}$ is maximal, i.e.
representing the global maximum of $\Delta \ln{Z}$. At $\beta =
\beta_{M}$ we have $\Delta \rho=0$, $\Delta \mathrm{Tr}(P)>0$
whereas just below the upper Hagedorn point
\begin{equation}
\Delta \mathrm{Tr}(P)(\beta, \phi_{m}(\beta)) < - \Delta
\rho(\beta, \phi_{m}(\beta)) \label{P}
\end{equation}
since $\Delta \mathrm{Tr}(P) = -3\Delta F$ when $\phi$ is constant
and $\ln{\Delta Z}$ will have a lower order zero at $\beta_{+}$
than its derivative w.r.t. $\beta$, so that for some $\beta_{C}$
with $\beta_{+} < \beta_{C} < \beta_{M}$ we have
$\epsilon(\beta_{C}, \phi_{m}(\beta_{C})=0$. These values give a
solution with $\phi$ and $g_{00}$ constant. Such a solution may be
seen as the interior of a gravitational conductor from which the
gravitational acceleration field has been expelled. The energy
density has a negative value and the pressure has a positive
value, both contributing to a negative spatial curvature: the
spatial geometry will be a Bolyai-Lobachevski 3-space with line
element
\begin{equation}
du^{2} =  \frac{dr^{2}}{1+a^{-2}r^{2}} + r^{2}(d \theta^{2} +
\sin^{2} \theta d \phi^{2}) \label{BL}
\end{equation}
giving space-time interval
\begin{equation}
ds^{2} =  dt^{2} - du^{2}
\end{equation}
with
\begin{equation}
a^{-2} = 8\pi G(p_{C}-\rho_{C}) \sim O(1)
\end{equation}
and $a^{-2}$ is positive since the constant values
$p_{C}$,$\rho_{C}$ for pressure and density satisfy $p_{C}>0$ and
$\rho_{C}<0$.
 This geometry provides an implementation of the holographic
principle [14,15], $\delta{V} \sim \delta{S}$. By eq.(\ref{DE}) we
have $\Delta S \sim -\Delta F > 0$ for the quasi-constant region
where $\Delta \epsilon \sim 0$.
 The excitations in this phase are strings with a higher
energy per unit length than free strings due to the condensate
$<\phi>$. The thermalon correlation length is given by
\begin{equation}
\lambda = \frac{1}{\mu_{\mathrm{eff}}}
\end{equation}
where $\mu_{\mathrm{eff}}$ is the effective thermalon mass. Now
because of spontaneous symmetry breaking, the thermalon in the BSP
is represented by two different modes, a Goldstone boson
$\phi_{G}$ with $\delta \phi_{G}/<\phi>$ imaginary, and a `Higgs'
mode $\phi_{H}$ with $\delta \phi_{G}/<\phi>$ real. The Goldstone
mode corresponds to a spectrum of strings at the critical
temperature, equivalent to free strings with
\begin{equation}
\gamma_{F}^{2}(G) = 6
\end{equation}
i.e. $\lambda(G) \rightarrow \infty$, whereas the Higgs mode gives
a spectrum which is equivalent to that for free strings at a value
of
\begin{equation}
\gamma_{F}^{2}(H)= 18-2\gamma^{2}
\end{equation}
for the quartic model.  As $\lambda(H)$ is shortest for $\beta =
\pi$ we can see that Higgs mode strings are relatively excluded
around the duality point, an effect which is essentially due to
the increased energy density of the strings. The effect of the
condensate $<\phi>$ on the string spectrum is equivalent to a
temporal dilatation effect. Strings propagate as if $g_{00}$ were
replaced by $g_{00}(G,H)$ with
\begin{equation}
g_{00}(G)=\frac{\beta_{F}(G)^{2}}{\beta^{2}}g_{00} =
\frac{\beta_{-}^{2}}{\beta^{2}}g_{00} \label{MG}
\end{equation}
and
\begin{equation}
g_{00}(H)= \frac{\beta_{F}(H)^{2}}{\beta^{2}}g_{00}, \label{MH}
\end{equation}
where $\beta_{F}(G,H)$ is the largest root of eq.(\ref{GAM}) for
$\gamma = \gamma_{F}(G,H)$. Another equivalent solution exists by
thermal duality by choosing the smaller roots for $\beta$, showing
that the physics is effectively continuous at both upper and lower
transition temperatures. The effective temporal dilatation
produced by $<\phi>$ applies as a Lorentzian boost to the
heterotic time lattice discussed above, preserving the self
duality of the lattice that is needed for consistent propagation
of the Goldstone and Higgs strings. The critical Goldstone strings
have neutral buoyancy due to the condensate, and are effectively
maintained at the Hagedorn transition throughout the BSP. As noted
above, the space-filling solution will have a temperature $T_{C}
=1/\beta_{C}$. Because of the effect of gravitational screening,
highly excited string states can exist within the BSP, whereas
gravity excludes most of them from the normal phase [4]. Thus the
greater entropy of the BSP can be understood as being due to the
gravitational screening within the BSP allowing modes such as
longer strings, which cannot exist within the normal phase because
of gravitational collapse. The next step is to examine the
possibility of solutions for finite regions of the high
temperature phase. This will allow a separation of space into
normal and BSP regions, with long strings effectively confined to
the latter. The idea is that the solution above would represent
limiting values for the interior region, and that a solution for a
finite extent would interpolate between this and an exterior
region of conventional vacuum, described by the same solution as
the exterior region of a black hole. At the surface where the
geometric temperature reaches the lower Hagedorn point, $\phi =
0$, and $\beta$ will interpolate inwards according to the
constraint eq.(\ref{VB}).

This condition gives the possibility of thermal equilibrium
between the interior region of broken symmetry and an external
vacuum region. How would this be possible without an event
horizon? In the Euclidean continuation, the region external to a
horizon can be described in radial coordinates. The radius $r$
represents distance from the horizon $r=0$, while imaginary time
wraps around circles of constant $r$ with period $\beta = 2\pi r$.
We envisage a solution where the interior tends to a constant
temperature, which would be equivalent to excising a disc around
$r=0$ and replacing this with a tube tending to a constant
circumference $\beta_{C}$. This tube would take us in past the
point where the horizon would have been, indeed right across the
interior region to open out to join another almost flat section at
the far boundary of the BSP. Thus a consistent geometric
temperature can be maintained without an event horizon because the
Euclidean continuation of the space-time is not simply connected.
The equilibrium will be typical of phases connected by a first
order phase transition coexisting at the transition temperature.
First consider a region with zero gauge charges. The degrees of
freedom within this surface will be $g_{ab}$, the spatial
components of the metric, $g_{00}$, the metric time component
($g_{\mu \nu}$ is the tensor string zero mode), and $\phi$. Some
configuration $C$ of these fields will give a global minimum of
\begin{equation}
-\ln{Z} = - \int \sqrt{det g_{ab}}d^{3}x(g_{4}^{-2}\beta R +
\Delta \ln{Z(\beta,\phi)}). \label{M}
\end{equation}
 For 4 dimensional string models where the
additional worldsheet degrees of freedom are subject to twisted
boundary conditions [2], there is no breathing mode dilaton
$\phi_{6}$, and the 4 dimensional dilaton $\phi_{4}$ decouples for
the heterotic string, hence they need not be included in
eq.(\ref{M}). (The suggestion in [4] that a modification of the
theory needs to be made to enable $\phi_{4}$ to decouple is
mistaken.) This feature gives a stable value for the gauge and
gravitational couplings, and indicates a special role for 4
dimensional heterotic models. The configuration $C$ will also be a
local minimum, so that it is a solution of the field equation
derived from eq.(\ref{M}). Consider the spherically symmetric
(zero angular momentum) case. At the centre, the partial
derivatives of the fields $\phi$ and $g$ all vanish. As the
components of $g$ will all be constrained by continuity with the
external values, there is essentially one degree of freedom: the
variation $\delta\phi$ of the thermalon field from its limiting
value, corresponding to different values of $M$, the total mass of
the region. Working inwards from the outer surface, the variations
of the fields from their limiting values decay exponentially over
a distance  of $O(\ln{M})$ to the centre, giving power law
dependences on $M$ there. Again this behaviour is suggestive of a
gravitational conductor, and there is a screening mass in the
boundary layer $O(1)$ per unit area, and $O(1)$ in thickness in
their dependence on $M$. For a region carrying gauge charges,
there will be a configuration which minimises
\begin{equation}
-\ln{Z} = - \int \sqrt{det g_{ab}}d^{3}x(g_{4}^{-2}\beta (R +
F^{2}) + \Delta \ln{Z(\beta,\phi)}), \label{ME}
\end{equation}
where $F$ represents the gauge fields, interpolating inwards from
a Reissner-Nordstrom type solution. It is likely that the gauge
charges will also be concentrated in the boundary layer of the
region. We anticipate that a generalisation to the rotating case
will also be possible, perhaps by giving a uniform angular
velocity to the interior. We may also note that although there
will be higher order corrections in $R$ and $F$ to eq.(\ref{ME}),
the mean value argument above should still apply. As  a further
investigation, for the simplest form of $\Delta \ln{Z}$ based on
eq.(\ref{QF}) an attempt may be made to solve the non-linear
differential equations generated by eqs.(\ref{M},\ref{ME}) either
numerically or in closed form, if this is possible.
\section{Nucleation Mechanism}
\ \ \ In this section we will consider how regions of the BSP
could form during gravitational collapse. The mechanism that we
propose is a kind of `diagravitational' breakdown of the normal
phase in globally strong gravitational fields. There are
significant differences with the dielectric breakdown of an
insulator in a strong electric field. One is the unexpected sign
of the effect (as like gravitational charges attract rather than
repel). The other is that, due to Einstein's equivalence
principle, there is no local trigger in the form of a strong local
invariant field strength. Rather, the transition must begin
simultaneously around a critical gravity surface. Such a surface
is found by the physical world `trying' all possible Feynman
paths, acting in this as a quantum-parallel computer. If such a
surface is found, the probability of nucleation rapidly switches
to almost $1$, so that the nucleation process acts as a non-local
`everywhere or nowhere' quantum switch. The long-distance
correlations of nucleation vs. non-nucleation would be of
Einstein-Podolsky-Rosen (EPR) type, and would not represent any
form of super-luminal causality. Consider a family of spacelike
surfaces $S(t)$ with $S^{2}$ topology connected by timelike curves
generated by a unit vector field $T$. Choose a frame field $E_{i}$
with $E_{0}= T$, $E_{2}$ and $E_{3}$ tangent to $S(t)$ (using two
or more patches) and $E_{1}$ an inward normal. Suppose that each
curve has acceleration $\alpha$ orthogonal to $S$ where the
associated inverse temperature $\beta = 2\pi/\alpha$ lies in the
range $\beta_{-} > \beta > \beta_{+}$. Applying the thermodynamic
equation $dS = \beta dQ$ and assuming for now the Hawking
relationship $dS = dA/4G$ (we will argue below that this is needed
for consistency) gives

\begin{equation}
d\dot{A} = T(dA) = d\dot{S}\frac{dA}{dS}=  \beta T_{01} \sigma
dA.4G = \frac{8\pi G}{\alpha}T_{01}\sigma dA
=\frac{\kappa}{\alpha}T_{01}\sigma dA   \label{A}
\end{equation}

where $0 \leq \sigma \leq 1$ is an opacity factor for the
absorbtion of energy at the surface and $\kappa$ is the constant
appearing in Einstein's field equations. Initially, we will have
$\sigma =0$, and we are led to consider surfaces with $d\dot{A}
=0$ and $\beta = \beta_{P}$ as nucleation sites, where we take
$\beta_{P} = \beta_{P}(\phi \sim 0)$. The polarization point value
for $\beta$ is chosen so that polarization may begin. For
notation, we will indicate a family with $d\dot{A}=0$ as
$S^{0}_{t}$ and if $\beta = \beta_{P}$ also, as
$S^{0}_{t}(\beta_{P})$. An element of a $S^{0}_{t}(\beta_{P})$
family may be regarded as a string regulated version of a Penrose
closed trapped surface [16], with such a Penrose surface being
obtained in the infinite tension limit.
  Now given such a family, a condensate $<\phi>
\neq 0$ around $S^{0}_{t}(\beta_{P})$ would produce gravitational
polarization with $\rho> 0$ outside $S^{0}_{t}(\beta_{P})$ and
$\rho < 0$ inside $S^{0}_{t}(\beta_{P})$. To produce such a
condensate requires a continuous phase for the condensate $<\phi>$
all around the spherical topology surfaces $S^{0}_{t}(\beta_{P})$.
This condensate would be self-sustaining as the outward
gravitational force on the inner part with $\beta < \pi$ would
balance the inward force on the outer part with $\beta
> \pi$, sustaining the acceleration relative to free-fall needed
to maintain the geometric temperature. Nucleation is now possible,
and with rising values of $\sigma$ the outer boundary of the
condensate follows eq.(\ref{A}), with the $\sigma$ value of the
condensate driven closer to $1$, while the inner boundary will be
at the innermost $S^{0}(\beta_{P})$ surface to have formed. The
existence of a $S^{0}_{t}(\beta_{P})$ family gives the coherence
needed to produce a stable BSP region. Because the boundary of the
BSP region is necessarily closed, the nucleation is a holistic
effect requiring the existence of such a family of closed
surfaces. Because the conditions defining a $S^{0}_{t}(\beta_{P})$
family apply locally (including continuity of $<\phi>$), in
computational terms the question of whether such a family exists
represents a non-local problem for which a candidate solution may
be confirmed by parallel local checking, and may be called a {\em
locally checkable problem \/}.

As the gravitational polarization progresses, pressure builds
within the shell. The spatial geometry is also significantly
affected by the energy needed to produce gravitational screening,
with the coefficient of $dr^{2}$ reaching the Bolyai-Lobachevsky
value $O(M^{-2})$ (see eq.(\ref{BL})) at the zero active gravity
(ZAG) point where $\epsilon$ vanishes. The average value of
curvature in the $r, \theta$ plane between the screen boundaries
and the ZAG surface is given by
\begin{equation}
R_{r, \theta} = \frac{\Delta \pi_{\mathrm{eff}}}{\pi \delta s}
\end{equation}
where $\delta s$ is the radial distance, and $\pi_{\mathrm{eff}}$
is the effective value of $\pi$ in the $r, \theta$ plane,
\begin{equation}
\pi_{\mathrm{eff}} = \frac{\delta r}{2\delta s}
\end{equation}
which is $(1-a^{-2}r^{2})^{1/2}$ in the BL region, or $\sim
O(1/M)$ for the boundaries of a static shell.
 This gives positive spatial curvatures of $O(1)$ in the
outer layers and negative spatial curvatures of $O(1)$ on the
inner layers in agreement with the magnitude and changing sign of
$\rho - p$, where $p$ is appropriate component of the pressure
matrix. The time needed to produce the radial deflation that
produces the BL geometry will be $O(\ln{M})$. From the viewpoint
of an observer at a constant area surface, the radial contraction
is initiated by the pulse of positive energy represented by the
ultra-relativistic in-falling matter, and then the radial metric
follows an exponential decay as the negative energy screen passes
through, stabilizing at the BL geometry. The formation of
gravitational screens may occur many times during the collapse
process, depending on details of the in-falling matter
distribution. Thus, in a local time $O(\ln{M})$ an outer
gravitational screen like the boundary of the BSP regions of
Section 2 will have formed, and the inner and outer gravitational
screens will separate, with a $\beta \sim \beta_{C}$ region
between them. The surface gravity in this region will be
effectively zero, rising to $2\pi/\beta_{P}$ on the inside and to
a somewhat smaller value $\sim 2\pi/ \beta_{-}$ on the outside,
giving a characteristic double peak pattern. The negative
gravitational energy formed at the inner part of the condensate
will cancel just enough of the energy of the in-falling matter
(which will be absorbed into the BSP) to maintain a value of
$2\pi/\beta_{P}$ as the maximum acceleration of $S^{0}$ surfaces,
so that the conversion of the interior to BSP proceeds by a
process of inward induction. Note that thermal equilibrium is
maintained at the outer surface of the BSP, but not at the inner
boundary which is intrinsically unstable, as the transition to the
BSP is in progress there. (Near the inner boundary $\Delta S$ may
become negative, suggesting that the inner screen may act as a
heat pump moving entropy to the $\Delta S > 0$ part of the BSP.)
The inner surface of the BSP will have $\beta < \beta_{C}$,
whereas the nascent $S^{0}_{t}(\beta_{P})$ surfaces have $\beta
=\beta_{P} > \beta_{C}$ and as the boundary moves further inward,
$\beta$ settles to $\beta_{C}$. Inward induction will be complete
when the whole of the interior is converted to a BSP region, which
will take a further time of at least $O(\ln{M})$ as seen from the
outer surface since the boundary between normal and BSP regions is
timelike.

During conversion to BSP, space from the normal phase will undergo
a radical radial contraction, and its contents will be
thermalised. In cases where there is a multiple concentric
nucleation of BSP regions, the negative energy inner screen of
each region will eventually encounter the positive energy outer
screen of the next region in, leading to coalescence. Eventually,
all matter will be absorbed and all regions of BSP will coalesce,
leading to the formation of a quasi-stable region. The time to
complete process of in-falling matter being absorbed at the outer
screen, causing this to expand outwards,  will be comparable to
the time taken for matter to fall to a critical gravity surface in
the Schwarzschild metric. This will be $O(\ln{M})$ as seen from
such a surface, or $O(M \ln{M})$ as seen from outside the strong
field region - see [17], section 15. This is comparable to the
conversion time for the interior by inward induction. The pressure
on the outer surface of the BSP region due to accretion will be
relatively small, at $O(1/\ln{M})$. A small amount of thermal
radiation will escape the very strong gravitational field of this
hot region, leading to a slow decay like that of a Hawking black
hole for an isolated BSP region. In a realistic astrophysical
context, further accretion would overwhelm this effect.  \

 \ \ We will next make some observations about the
relationship of the nucleation process to the framework of quantum
mechanics. The appearance of large closed surfaces as nucleation
sites is an indication of that nucleation is a non-local quantum
process. There does not appear to be a problem with superluminal
transmission of information, as a strong gravitational field on
the same spatial scale is required to provoke a phase transition,
and this would take time $O(M)$ to set up, and the initial
nucleation process takes time $O(M)$ as seen from the outside .
The conversion of a collapsing object to a BSP region would be
analogous to the operation of a quantum information processing
device. If we separate space into an inner region $I$ and outer
region $I'$ by a closed surface $S = \delta I$, a quantum state
may be represented as
\begin{equation}
|\psi> = |\psi,I'>^{i}|I>_{i}.
\end{equation}
The Hilbert space is decomposed as $H = H_{I'} \otimes H_{I}$ and
the states $|I>_{i}$ form a basis for $H_{I}$ while the
coefficient states $|\psi,I'>^{i}$ lie in $H_{I'}$. Here the index
$i$ is specific to $S$ (an abbreviation for $i(S)$). If our
knowledge of $|\psi>$ is restricted to $I$, or we have no control
over the relative phases of the coefficient states, this may
represented by the density matrix
\begin{equation}
\rho^{j*,i} = <\psi,I'|\psi,I'>^{j*,i}
\end{equation}
where the index $j*$ refers to $<\psi,I'|$, or
\begin{equation}
\rho =<\psi,I'|\psi,I'>^{j*,i}|I>_{i}<I|_{j*},
\end{equation}
giving an entropy of entanglement $S = Tr \rho \ln{\rho}$. Now the
relationship of entanglement between $I$ and $I'$ is mutual, and
reversing the roles of $I,I'$ in the above will give a density
matrix $\rho '$ over $I$ giving the same value of $S$. The proof
is easy: if $\mathrm{dim}(H_{I})=N$, then the $|\psi,I'>^{i}$ span
a space $V$ of dimension $K \leq N$. Let $|W>_{k}$ be an
orthonormal basis of $V$ made up, if necessary, to $N$ elements.
Then we can write
\begin{equation}
|\psi> = a^{k,i}|W>_{k}|I>_{i}.
\end{equation}
Now $\rho =a^{\dag}a$ while $\rho' =aa^{\dag}$,and both matrices
are evidently Hermitian. The spectra of eigenvalues of $\rho$ and
$\rho'$ will be the same, since
$Tr(a^{\dag}a)^{n}=Tr(aa^{\dag})^{n}$ for any $n$. In fact, if we
choose the bases $|I>_{i}$ and $|W>_{k}$ to consist of
eigenvectors of $\rho$ and $\rho'$ respectively, then $\rho =
\rho'$. (If the entanglement between $I$ and $I'$ is produced by a
measurement process, these bases represent the eigenstates of the
measured quantities.) The equality of the entropies calculated
from $\rho$ and $\rho'$ follows. Thus $S$ also represents the
entropy of ignorance of an observer in $I'$ who has no control
over the phases of coefficient states within $I$. We will now
suppose a form of the holographic principle, which is that the
entropy of entanglement for any state $|\psi>$ between $I$ and
$I'$ is bounded by $\sim \exp{\kappa A(S)}$ for some constant
$\kappa$. The two special cases that we need to consider to
generate a calculation for the general case are that $I$
represents (1) a region of the normal phase and (2) a region of
BSP. In case (1), the entropy bound is supported by the analysis
of Srednicki [18]. The following (very) heuristic argument points
in the same direction. To measure and record information $U$
within $I$ by generating entanglements with uncontrolled phases to
the environment $I'$ requires energy $E \geq U / \beta$ at inverse
temperature $\beta$. (At large $\beta$ there will be a shortage of
low energy modes within $I$ and an inequality will definitely be
needed.) Now at $S$, $E$ will produce an average gravitational
acceleration of $4 \pi GE/A(S)$ giving a geometric $\beta =
A(S)/2GE$, and using this to solve $U \leq \beta E$ gives $U \leq
A(S)/2G$. Presumably, a careful treatment would give consistency
with the Hawking value $A(S)/4G$ for black holes. For case (1) it
seems that the `strong' holgraphic principle holds, i.e
$\mathrm{dim}(H_{I})$ is bounded by a multiple of $A(S)$ as highly
excited states are liable to gravitational collapse. Thus, for
example, very high excitation levels of bosonic field modes within
$I$ would be cut off by the transition to the BSP.  For case (2),
the combination of hyperbolic spatial geometry and the lower bound
on $\gamma$ give a finite upper bound on $\Delta S$ consistent
with the `weak' holographic principle, even though gravitational
screening means that there is no further collapse and so the
dimension of $H_{I}$ is not necessarily finite in this case. The
contribution of the Goldstone strings is understood to be included
within $\Delta S$.

\ \ \ For concentric regions $I,J$ with $ I \supset J$, $R = I-J$
we may write
\begin{equation}
|I>_{i} = |R>^{j}_{i}|J>_{j}
\end{equation}
so that
\begin{equation}
|\psi> = |\psi,I'>^{i}|I>_{i} = |\psi,I'>^{i}|R>^{j}_{i}|J>_{j}.
\end{equation}
Here  $H_{I}$ is decomposed as $H_{R} \otimes H_{J}$ with
$|R>^{j}_{i}$ in $H_{R}$. $H_{R}$ may be viewed as a kind of
quotient space with
\begin{equation}
\mathrm{dim} (H_{R}) = \frac{\mathrm{dim} (H_{I})}{\mathrm{dim}
(H_{J})}.
\end{equation}
Resolving gives
\begin{equation}
|R>^{j}_{i} = C^{j,r}_{i}|R>_{r}
\end{equation}
where the combined index $j,r$ takes as many values as $i$. Thus
\begin{equation}
|\psi> = |\psi,I'>^{i}C^{j,r}_{i}|R>_{r}|J>_{j}.
\end{equation}
 During the transition we may take $R$ to be the region of BSP, bounded on
the outside by $S_{+}$ and on the inside by $S_{-}$. Now
$\exp({\int_{R}\Delta \ln{Z} dv})/Z(R)_{\mathrm{infalling}}$
represents the relative probability of nucleation over no
nucleation ($dv$ is the 3-dimensional BSP volume element), and as
$\int_{R} \Delta \ln{Z} dv \sim O(A(S_{+})-A(S_{-}))$, this will
usually be overwhelming in gravitational collapse where the
conditions for nucleation are met. Thus we have an automatic
inter-phase conversion process of the states in $R$
\begin{equation}
M_{\phi}: |0,R>_{r}  \rightarrow |<\phi>,R>_{r}
\end{equation}
where the states $|<\phi>,R>_{r}$ have the condensate $<\phi>$,
the states on the left having $<\phi> = 0$. As noted above, these
states will an effective degeneracy of $\exp({\int_{R} \Delta S
dv})$. What are the `microstates' being counted by this formula?
The answer to this question is provided by the Goldstone string
modes of the BSP. They provide a heat bath giving a zero
contribution to $\Delta \ln{Z}$ but a positive contribution to
$\Delta S$, and can adjust the contribution of the Higgs strings
to give the correct value for $\Delta S$. Because the states of
the normal phase are converted rather than copied, the transition
process does not conflict with the `Xerox principle' [6]. Note
that because
\begin{equation}
\frac{\partial \beta(G)}{\partial \beta} =0
\end{equation}
the Goldstone strings have zero effective gravitational energy,
(i.e. they displace an equal amount of energy from the condensate)
and because there is no gravitational gradient for them, there is
no build up of pressure, and their contribution to the
gravitational source is therefore zero. The absence of pressure
due to the Goldstone strings indicates that this component of the
BSP behaves more like a single long string than a gas of short
strings. Introducing Goldstone string excitations to reconcile
$\Delta S$ thus does not modify the spacetime metric, while any
modification due to Higgs strings should not alter the qualitative
features of the metric. The negative value of $\Delta \ln{Z}$ is
provided by the symmetry breaking phase transition (the potential
energy of $<\phi>$), while $\Delta S$ is provided by the extra
string modes of the BSP. At the end of the nucleation process,
when $R$ no longer has an inner boundary, $R=I$ and since this
region is in thermal equilibrium with the surrounding vacuum,
consistency seems to require
\begin{equation}
\int_{I} \Delta S dv = S_{\mathrm{Hawking}} \label{H}
\end{equation}
as information can be freely exchanged between the two phases. The
effective potential for $\phi$ should therefore be consistent with
eq.(\ref{H}). The quantity $\Delta S$ may be divided into a `bulk'
component $\Delta S_{\mathrm{int}}$ where $\beta \sim \beta_{C}$
and a `surface' boundary layer component $\Delta
S_{\mathrm{surf}}$. Due to the hyperbolic geometry of the
interior, both are proportional to the surface area. We would
expect that $\Delta S/\Delta A$ would be less for the surface
components (i.e the average for the outer and inner screens) than
for the bulk, so that the coalescence process for BSP shells would
be thermodynamically irreversible (i.e. the reverse process would
be very unlikely). Due to the onset of hyperbolic geometry, the
inner area of the surface component would be a fraction $0 < r <
1$ of the outer area, so the quantity $\Delta S/\Delta A$ would
tend to a constant value for large $A$, so that this condition
would be reasonable. The two phases would coexist locally at the
Hagedorn temperature, with the latent heat of the transition used
to produce more Goldstone and Higgs strings. These would act as an
inner heat bath in equilibrium with the external Unruh/Hawking
radiation (except for the small rate of escape). Thus the
nucleation mechanism acts as a surface gravity regulator, and is
the mechanism which implements the gravitational bound on the
information content of the normal phase. If we try to force too
much information into a region $I$, the gravitational field of the
energy used to store this information will cause BSP nucleation
from the boundary surface $S$. Because the holographic principle
applies in both the normal and BSP cases it is possible for
$M_{\phi}$ to be one-to-one so that the information content of a
collapsing body can be preserved. It is also worth noting that the
outer gravitational screen will have states that are strongly
entangled with the states of the inner gravitational screen and,
as the inner screen rolls inwards, with the interior of $R$, so
that it will also act as a thermal information screen, or quantum
encryption device, in analogy to the event horizon of a Hawking
black hole. Thus $\int_{I}\Delta S dv$ will also represent the
apparent change of entropy to an observer who has access only to
$I'$. The BSP will act as a heat bath for the surrounding space,
much as the horizon does in Hawking's theory. As in Hawking's
theory, a small amount of radiation will escape the strong
gravitational field of the BSP. During the collapse and radiative
decay, the BSP acts as an information bank to hold quantum
information about the collapsed object. The nucleation process
acts as an extremely sensitive {\em detector \/} which captures
essentially all details of the local environment and records them
in a region of the quasi 2-dimensional BSP. This process could
therefore be viewed as a form of {\em flash holography \/},
converting the collapsing region to a holgram of itself, given the
rapid and non-local nature of the transition.

\section{Conclusions}
\ If the scenario presented in this paper turns out to be viable,
some of the unsatisfactory aspects of the physics of black holes
may be avoided. These include the presence of singularities and
the problem of time reversibility if black holes are allowed but
their time reversals (white holes) are not. The time asymmetry in
forming BSP regions is statistical only, and there is no clash
with the CPT theorem. The formation of quasi two-dimensional BSP
regions can be viewed as a real hologram formation mechanism,
preserving the information content of the collapsed body. Finally
we may note an analogy with the physics of the strong interaction,
with long string modes confined within BSP regions, where gravity
is effectively weak, much as quarks are confined within hadrons
where they are asymptotically free.\

\

The author would be very grateful for any comments or suggestions
on the contents of this paper.

\

\

\

\textbf{References} \

 \

[1] D.Gross, J.Harvey, E.Martinec and R.Rohm,
 {\em Nucl. Phys. B\/}{\bf 256} (1985) 253;
 {\bf 267} (1986) 75.

[2] H.Kawai, D.C.Lewellen and S.H.H.Tye, {\em Nucl. Phys. B\/}{\bf
288} (1987) 1.

[3] J.Hartle and S.W.Hawking, {\em Phys. Rev. D\/}{\bf 13} (1976)
2108.

[4] M.Hewitt, {\em Phys. Lett. B\/}{\bf 309} (1993) 264.

[5] L.Susskind, {\em hep-th\/}/9309145.

[6] D.Bigatti and L.Susskind {\em hep-th\/}/0002044.

[7] R.Hagedorn, {\em Nuovo Cim.\/}{\bf 64} A (1965).

[8] B.Sathiapalan, {\em Phys. Rev. D\/}{\bf 35} (1987) 3277.

[9] K.H.O'Brien and C.I.Tan, {\em Phys. Rev. D\/}{\bf 36} (1987)
1184.

[10] J.J.Atick and E.Witten, {\em Nucl. Phys. B\/}{\bf 310} (1988)
291.

[11] W.G.Unruh, {\em Phys. Rev. D\/}{\bf 14} (1976) 870.

[12] N.D.Birrell and P.C.W.Davies, {\em Quantum fields in curved
space\/} (CUP, Cambridge 1982).

[13] R.Rohm, {\em Nucl. Phys. B\/}{\bf 237} (1984) 553.

[14] L.Susskind, {\em hep-th\/}/9409089.

[15] C.R.Stephens, G.'t Hooft and B.F.Whiting, {\em
gr-qc\/}/9310006.

[16] R.Penrose, {\em Phys. Rev. Lett.\/}{\bf 14}(1965) 57.

[17] A.P.Lightman, W.H.Press, R.H.Price and S.A.Teukolsky, {\em
Problem book in relativity and gravitation\/} (Princeton
University Press, Princeton 1975)

[18] M.Srednicki, {\em Phys.Rev.Lett\/}{\bf 71} (1993) 666.

\end{document}